\documentclass[nofootinbib,prd]{revtex4}%
\usepackage{amsmath}
\usepackage{amsfonts}
\usepackage{amssymb}
\usepackage{graphicx}%
\usepackage{amsmath,amssymb}
\usepackage{mathrsfs}
\usepackage{graphicx}
\usepackage{color}
\usepackage{subfigure}
\usepackage{fancyhdr}
\usepackage{multirow}
\usepackage{float}
\usepackage{epsfig}
\usepackage{amsfonts}
\usepackage{bm}

\begin{document}
\title{Hot Casimir Wormholes}

\author{Remo Garattini}
\email{remo.garattini@unibg.it}
\affiliation{Universit\`a degli Studi di Bergamo, Dipartimento di Ingegneria e Scienze
Applicate, Viale Marconi 5, 24044 Dalmine (Bergamo) Italy and I.N.F.N.-
sezione di Milano, Milan, Italy.\\Canadian Quantum Research Center 204-3002 32 Ave Vernon, BC V1T 2L7 Canada}
\author{Mir Faizal}
\email{mirfaizalmir@gmail.com}
\affiliation{Canadian Quantum Research Center 204-3002 32 Ave Vernon, BC V1T 2L7 Canada, \\Irving K. Barber School of Arts and Sciences,\\ University of British Columbia Okanagan, Kelowna, BC V1V 1V7, Canada.}

\vspace{-0.1cm}

\begin{abstract}
    In this paper, we have for the first time considered the consequences of thermal fluctuations to the Casimir effect on a traversable wormhole. This was done by using finite temperature generalization of the Casimir effect as a source of a hot traversable wormhole. Thus, we have considered a more physical scenario, where the effects of thermal fluctuations are also considered as a source of a traversable wormhole.  To obtain a dependence on such a thermal Casimir effect,  consider the plates positioned at a distance either parametrically fixed or radially varying. In both cases, the temperature effects are investigated.  We demonstrate that thermal fluctuations modify the throat of the wormhole. Such results have been obtained in both regimes, i.e. high temperature and low temperature. We explicitly investigate the effect of such finite temperature effects on the size of a wormhole. 
\end{abstract}
\maketitle
\section{Introduction}
A  wormhole is a solution of the Einstein Field Equations (EFE) that acts as a tunnel from one region of spacetime to another. The first wormhole solutions considered were unstable, and it was not possible for observers to pass through them \cite{0, 1}. Unlike these unstable wormholes, it is possible to also construct stable wormhole solutions, such that observers can pass through them, and this is done by violating the weak energy condition  \cite{c, c1, c2}.  Such a stable wormhole solution is called a Traversable Wormhole (TW).   As classical solutions obey the weak energy conditions, the construction of TW has to crucially depend on quantum effects, which can violate the weak energy conditions \cite{ MT, MTY,Visser,cw12}. 
A substance where the quantum effects violate weak energy conditions is known as exotic matter \cite{d, d1}. It is known that vacuum energy produced from quantum fluctuations critically depends on purely quantum effects, so it can act as exotic matter. In fact, it is also known that these quantum fluctuations produce an effective force on the macroscopic boundaries made by conducting surfaces, and this effect is called the Casimir effect \cite{Casimir}. 
It has been explicitly demonstrated that the Casimir effect does violate weak energy conditions  \cite{w1, w2}. So,  the Casimir effect can act as a form of exotic matter and source TW, and such TW are also called Casimir wormholes \cite{CW}.

Now at zero temperature, the Casimir effect between two plane parallel, closely
spaced, uncharged, metallic plates in a vacuum. predicts an attractive force of
the form
\begin{equation}
F(d)  =-\frac{3{\hbar c\pi^{2}}S}{720d^{4}},
\end{equation}

where $S$ is the surface of the plates and $d$ is the separation between them  \cite{Casimir}.
The force $F\left(  d\right)  $ is also responsible to produce a pressure
\begin{equation}
P(d)  =\frac{F(d)}{S} =-\frac{3{\hbar c\pi^{2}}}{720d^{4}}.
\end{equation}

Both of them are obtained with the help of the renormalized energy
\begin{equation}
E^{\text{Ren}}(d)  =-\frac{\hbar c\pi^{2}S}{720d^{3}}.
\end{equation}

It is immediate to recognize that the energy density is obtained by dividing
$E^{\text{Ren}}( d)  $ by the volume $V=Sd$, yielding
\begin{equation}
\rho_{C}\left(  d\right)  =-\frac{\hbar c\pi^{2}}{720d^{4}}.  
\end{equation}

We can see that a relationship exists between the energy density $\rho
_{C}\left(  d\right)  $ and the pressure $P\left(  d\right)  $. This is
described by an Equation of State (EoS) of the form $P=\omega\rho$ with
$\omega=3$. As far
as we know, the Casimir energy represents the only artificial source of
\textit{exotic matter}\ realizable in a laboratory\footnote{Actually, there
exists also the possibility of taking under consideration a squeezed vacuum.
See for example Ref.\cite{Hochberg}.}\textit{.} Exotic matter violates the
Null Energy Condition (NEC), namely for any null vector $k^{\mu}$, we have
$T_{\mu\nu}k^{\mu}k^{\nu}\geq0$. Such exotic matter can produce TW solutions to the  EFE  \cite{cw12, MT, MTY}. It may be noted that the only quantum effects in this case are needed to produce the exotic matter, and the rest of the treatment remains classical  \cite{MT,Visser}.
As the geometry remains classical, the EFE must be
replaced with the semiclassical EFE, namely
 $
G_{\mu\nu}=\kappa\left\langle T_{\mu\nu}\right\rangle ^{\text{Ren}}%
 $and $ \kappa={8\pi G}/{c^{4}},
$ 
where $\left\langle T_{\mu\nu}\right\rangle ^{\text{Ren}}$ describes the
renormalized quantum contribution of some matter fields. To further proceed, we introduce the
following spacetime metric%
\begin{equation}
ds^{2}=-e^{2\Phi\!\left(  r\right)  }\,dt^{2}+\frac{dr^{2}}{1-b(r)/r}%
+r^{2}\,(d\theta^{2}+\sin^{2}{\theta}\,d\varphi^{2})\,, \label{metric}%
\end{equation}
representing a spherically symmetric and static wormhole. $\Phi\!\left(
r\right)  $ and $b(r)$ are arbitrary functions of the radial coordinate
$r\in\left[  r_{0},+\infty\right)  $, denoted as the redshift function, and
the shape function, respectively \cite{MT,Visser}. A fundamental property of a
a wormhole is flaring out of the condition of the throat, given by
$(b-b^{\prime}r)/b^{2}>0$, must be satisfied as well as the request that
$1-b(r)/r>0$. Furthermore, at the throat $b(r_{0})=r_{0}$ and the condition
$b^{\prime}(r_{0})<1$ is imposed to have wormhole solutions. It is also
fundamental that there are no horizons present, which are identified as the
surfaces with $e^{2\Phi\!\left(  r\right)  }\rightarrow0$, so that
$\Phi\!\left(  r\right)  $ must be finite everywhere. With the help of the
line element $\left(  \ref{metric}\right)  $, we can write the EFE in an
orthonormal reference frame, leading to the following set of equations%
\begin{equation}
\frac{b^{\prime}\left(  r\right)  }{r^{2}}=\kappa\rho\left(  r\right)  ,
\label{rho}%
\end{equation}%
\begin{equation}
\frac{2}{r}\left(  1-\frac{b\left(  r\right)  }{r}\right)  \Phi\!^{\prime
}\left(  r\right)  -\frac{b\left(  r\right)  }{r^{3}}=\kappa p_{r}\left(
r\right)  , \label{pr}%
\end{equation}%
\begin{align}
&  \Bigg\{\left(  1-\frac{b\left(  r\right)  }{r}\right)  \left[  \Phi
^{\prime\prime}\!\left(  r\right)  +\Phi\!^{\prime}\left(  r\right)  \left(
\Phi^{\prime}\!\left(  r\right)  +\frac{1}{r}\right)  \right] \nonumber\\
&  -\frac{b^{\prime}\left(  r\right)  r-b\left(  r\right)  }{2r^{2}}\left(
\Phi\!^{\prime}\left(  r\right)  +\frac{1}{r}\right)  \Bigg\}=\kappa p_{t}(r),
\label{pt}%
\end{align}
in which $\rho\left(  r\right)  $ is the energy density\footnote{However, if
$\rho\left(  r\right)  $ represents the mass density, then we have to replace
$\rho\left(  r\right)  $ with $\rho\left(  r\right)  c^{2}.$}, $p_{r}\left(
r\right)  $ is the radial pressure, and $p_{t}\left(  r\right)  $ is the
lateral pressure. We can complete the EFE with the expression of the
conservation of the stress-energy tensor which can be written in the same
orthonormal reference frame
\begin{equation}
p_{r}^{\prime}\left(  r\right)  =\frac{2}{r}\left(  p_{t}\left(  r\right)
-p_{r}\left(  r\right)  \right)  -\left(  \rho\left(  r\right)  +p_{r}\left(
r\right)  \right)  \Phi\!^{\prime}\left(  r\right).
\end{equation}

 An investigation on the connection between a TW and the Casimir energy has been done \cite{MTY}. Furthermore,  it has been examined which kind of TW one can obtain with a
Casimir source \cite{CW}. This kind of special Traversable Wormhole has been dubbed Casimir Wormhole.  The 
electromagnetic field combined with the Casimir source has been used to study the effects of
such an electrovacuum on a TW  \cite{CCW}.  The TW produced by a  Casimir source, with a scalar field has been studied both  with and without
a potential term \cite{SCW}. The  Einstein-Maxwell theory coupled to charged massless fermions, has also been used to construct TWs  \cite{cw12}. It has been argued that such solutions can be embedded in the standard model by making the overall size of the wormhole smaller than the electroweak scale. This solution is viewed as a pair of entangled near-extremal black holes, and there is an interaction term generated by the fermionic fields. It is known that wormholes can be viewed as a pair of entangled black holes \cite{b1, b2}. This has even motivated the study of traversable wormholes as quantum processors \cite{cw14}. This can be done by analyzing the holographic dual to a traversable wormhole. Specifically, the holographic dual has been analyzed using the  SYK many-body system. 

The Casimir wormholes in modified theories of gravity have also been studied. The modifications to such wormholes in gravity with higher scalar curvature and torsion terms have been studied \cite{ft, ft01}. The effect of noncommutativity on Casimir wormholes in higher dimensional Gauss–Bonnet gravity has also been studied \cite{gb}. It was observed that both noncommutativity and higher curvature terms modify the behavior of Casimir wormholes.
It is possible to analyze low energy collective excitations of a magnetically charged black hole using a large number of Alfven wave modes \cite{phys1}. The  Casimir energy of the Alfven wave modes has been used to construct a traversable wormhole \cite{phys}. 
It is possible to generalize the usual Casimir effect to a non-abelian Casimir effect, which is produced by  Yang-Mills theory \cite{ym}. This non-abelian  Casimir effect has also been used to construct traversable wormhole solutions \cite{nb}. The effect of the minimal length on Casimir wormholes has also been considered \cite{gup1, gup2, gup3}. Such minimal length occurs due to quantum gravitational modifications to the low energy quantum mechanics \cite{un} and quantum field theory \cite{un1, un2}. The generalization of Casimir wormholes to gravity's rainbow has also been studied \cite{gr}. In gravity's rainbow, the spacetime geometry depends on the energy of the probe, and this energy dependence can modify the behavior of the wormhole. Thus, the modification of gravity does modify the behavior of Casimir wormholes. However, in all these cases, the magnitude of the Casimir effect remains small, and so the size of the wormhole also remains small. This is because the wormhole in all these modifications to gravity is still produced by quantum fluctuations, and their magnitude at large scales remains small.

The Casimir effect assumes perfectly conducting surfaces. However, actual plates,   are never perfectly conducting. Rather they are characterized by a complex permittivity, with a material-dependent function of frequency. This consideration modifies the original zero-temperature Casimir force, to a finite-temperature thermal effect. In this finite temperature effect,  the effects of thermal fluctuations are considered besides the effects produced by quantum fluctuations.
This can be done using the Lifshitz theory, in which the electromagnetic field stress tensor is obtained by considering the correlated fluctuating charges and currents in the plates \cite{li}.
This finite temperature generalization of the conventional Casimir effect is called the thermal Casimir effect \cite{th1, th2}. It has been observed that the force produced due to thermal fluctuations dominates over the force produced due to zero temperature quantum fluctuation at separations greater than a critical value \cite{na, na2}. 
Thus, it seems possible that they could become important in scaling up the size of TWs. In fact, a serious problem with using the Casimir effect as a source of exotic matter is that it can only produce very small TWs, as the Casimir effect is very small.  However, it is both important and interesting to scale up the size of such wormholes. 
Even though it seems natural to consider the thermal Casimir effect for this, this has never been considered. 
Hence, we for the first time study the effect of temperature on TW and construct hot Casimir wormholes. The rest of the paper is structured as follows: in section \ref{p2}, we examine the low-temperature contribution to the Casimir energy source for a Traversable wormhole with the plates
positioned at a distance either parametrically fixed or radially varying. In
section \ref{p3}, we examine the high-temperature contribution to the Casimir
energy source for a Traversable wormhole with the plates positioned at a distance either
parametrically fixed or radially varying. We summarize and conclude in section
\ref{p4}.

\section{Low-Temperature corrections to the Casimir Wormhole}

\label{p2}
 At finite temperature, the thermal Casimir effect predicts the
following corrections on the free energy at low temperature%
\begin{equation}
\qquad E\left(  d,T\right)  =-\frac{\hbar c\pi^{2}S}{720d^{3}}\left[
1+\frac{45\zeta\left(  3\right)  }{\pi^{3}}\left(  \frac{T}{T_{eff}}\right)
^{3}-\left(  \frac{T}{T_{eff}}\right)  ^{4}\right]  ;\qquad T\ll T_{eff},
\label{ELow}%
\end{equation}

where we have defined%
\begin{equation}
T_{eff}=\frac{\hbar c}{2dk_{B}}. \label{Teff}%
\end{equation}
To have an order of magnitude on $T_{eff}$, we can put some numbers and we get%
\begin{equation}
T_{eff}=\frac{\hbar c}{2k_{B}d}\simeq\frac{\left(  10^{-34}\mathrm{J\ s}%
\right)  \left(  10^{8}\mathrm{m\ s}^{-1}\right)  }{2d\left(  10^{-23}%
\mathrm{J\ K}^{-1}\right)  }\simeq\frac{10^{-3}}{d}m\ K.
\end{equation}
Therefore for a plates separation of the order of $10^{-6}m$, we find that%
\begin{equation}
T_{eff}\simeq5\times10^{2}K.
\end{equation}
From Eqs. $\left(  \ref{ELow}\right)  $ and $\left(  \ref{EHigh}\right)  $, it
is possible to compute the related pressures. Note that the cubic term is
independent on $d$ in Eq.$\left(  \ref{ELow}\right)  $. For low temperatures,
we find%
\begin{equation}
P\left(  d,T\right)  =-3\frac{\hbar c\pi^{2}}{720d^{4}}\left[  1+\frac{1}%
{3}\left(  \frac{T}{T_{eff}}\right)  ^{4}\right]  \qquad T\ll T_{eff}.
\label{PL}%
\end{equation}
Eq.$\left(  \ref{ELow}\right)  $ will be the cornerstone of this
section. Following Ref.\cite{CW} we can consider the plates positioned at a
distance either parametrically fixed or radially varying. We have two possible configurations:

\begin{enumerate}
\item We divide $E\left(  d,T\right)  $ with a volume term of the form $V=Sd$,
leading to the following form of the energy density%
\begin{equation}
\qquad\rho_{L,1}\left(  d,T\right)  =-\frac{\hbar c\pi^{2}}{720d^{4}}f\left(
T,d\right)  , \label{rhoL1}%
\end{equation}
where we have introduced the following temperature function%
\begin{equation}
f\left(  T,d\right)  =1+\frac{45\zeta\left(  3\right)  }{\pi^{3}}\left(
\frac{T}{T_{eff}}\right)  ^{3}-\left(  \frac{T}{T_{eff}}\right)  ^{4}.
\label{f(T)}%
\end{equation}

\item Following Ref. \cite{CW}, we promote the plates distance $d$ as a radial
variable $r$ and we divide $E\left(  r,T\right)  $ with a volume term of the
form $V=Sr$, leading to the following form of the energy density%
\begin{equation}
\qquad\rho_{L,2}\left(  r,T\right)  =-\frac{\hbar c\pi^{2}}{720r^{4}}f\left(
T,r\right)  , \label{rhoL2}%
\end{equation}
where we have substituted the plates distance $d$ with the radial variable $r$
into the Eq.$\left(  \ref{f(T)}\right)  $. Because of thermal corrections,
this time the EoS between $\rho_{C}\left(  d,T\right)  $ and the pressure
$P\left(  d,T\right)  $ changes. Indeed, for low temperatures, we find
$P_{L}\left(  d,T\right)  =\omega_{L}\left(  d,T\right)  \rho_{L}\left(
d,T\right)  $ with%
\begin{equation}
\omega_{L}\left(  d,T\right)  =\frac{3}{f\left(  T,d\right)  }\left(
1+\frac{1}{3}\left(  \frac{T}{T_{eff}}\right)  ^{4}\right)  . \label{omegaL}%
\end{equation}
This relationship is true even when we substitute $d$ with $r$. We begin to
consider the configuration 1)
\end{enumerate}

\subsection{Constant Plates distance separation at Low Temperature}

If we give a look at the energy density $\left(  \ref{rhoL1}\right)  $, we can
see that this is constant with respect to the radial variable $r$. Therefore
the first EFE leads to the following shape function
\begin{equation}
b\left(  r\right)  =r_{0}-\frac{8\pi G}{c^{4}}\left(  \frac{\hbar c\pi^{2}%
}{720d^{4}}f\left(  T,d\right)  \right)  \left(  r^{3}-r_{0}^{3}\right)
=r_{0}-\frac{r_{1}^{2}\left(  T\right)  }{3d^{4}}\left(  r^{3}-r_{0}%
^{3}\right)  ,\label{b(r)LTd}%
\end{equation}
where we have defined%
\begin{equation}
r_{1}^{2}\left(  T\right)  =\frac{\hbar G}{c^{3}}\frac{\pi^{3}}{90}f\left(
T,d\right)  =\frac{\pi^{3}\ell_{P}^{2}}{90}f\left(  T,d\right)  =r_{1}%
^{2}\,f\left(  T,d\right)  .
\end{equation}
Even if the flare-out condition%
\begin{equation}
b^{\prime}\left(  r_{0}\right)  =-\frac{r_{0}^{2}r_{1}^{2}\left(  T\right)
}{d^{4}}<1
\end{equation}
is always satisfied, the shape function $\left(  \ref{b(r)LTd}\right)  $ does
not represent strictly a TW, because it is not asymptotically flat. However, we can observe
that there exists%
\begin{equation}
\bar{r}=r_{0}\sqrt[3]{1+\frac{3d^{4}}{r_{1}^{2}\left(  T\right)  r_{0}^{2}}}%
\end{equation}
such that $b\left(  \bar{r}\right)  =0$. Since $T\ll T_{eff}$, from
Eq.$\left(  \ref{f(T)}\right)  $, we can restrict the range of $f\left(
T\right)  $ to the values $\left[  1,1.12\right]  $. At $T=0$, the energy
density $\left(  \ref{rhoL1}\right)  $ has been taken under consideration in
the picture of a Generalized Absurdly Benign Traversable Wormhole
(GABTW)\cite{GABTW}. Nevertheless, we can see that there exist other
possible configurations generated by the shape function $\left(  \ref{b(r)LTd}\right)
$. Indeed, plugging Eq.$\left(  \ref{b(r)LTd}\right)  $ into Eq.$\left(
\ref{pr}\right)  $, one finds%
\begin{equation}
\Phi'\!\left(  r\right)  =\frac{\left(  r_{0}^{3}-\left(
3\omega_{L}+1\right)  r^{3}\right)  r_{1}^{2}\left(  T\right)  +3r_{0}d^{4}%
}{\left(  2r^{4}-2rr_{0}^{3}\right)  r_{1}^{2}\left(  T\right)  +6r\,d^{4}%
\left(  r-r_{0}\right)  },\label{PhiLTd}%
\end{equation}
where we have used the EoS $P_{L}\left(  d,T\right)  =\omega_{L}\rho
_{L,1}\left(  d,T\right)  $. Close to the throat, one finds%
\begin{equation}
\Phi'\!\left(  r\right)  \simeq\frac{r_{0}^{2}\left(  1-\left(
3\omega_{L}+1\right)  \right)  r_{1}^{2}\left(  T\right)  +3d^{4}}%
{6d^{4}\left(  r-r_{0}\right)  }.
\end{equation}
In order to avoid a horizon, we assume that
\begin{equation}
\omega_{L}=\frac{d^{4}}{r_{0}^{2}r_{1}^{2}\left(  T\right)  }.\label{oL}%
\end{equation}
Comparing Eq.$\left(  \ref{omegaL}\right)  $ and Eq.$\left(  \ref{oL}\right)
$ we also obtain%
\begin{equation}
\qquad r_{0}=r_{0}\left(  T,d\right)  =\frac{d^{2}}{\sqrt{3\left(  1+\frac
{1}{3}\left(  \frac{T}{T_{eff}}\right)  ^{4}\right)  }r_{1}},\label{r0L}%
\end{equation}
namely the size of the throat has a dependence on the temperature $T$.
However, since $T\ll T_{eff}$, the correction with respect the zero
temperature case is negligible. From Eq. $\left(  \ref{r0L}\right)  $, it is
possible to estimate the boundary location. Indeed, one finds
\begin{equation}
\bar{r}=\frac{d^{2}}{\sqrt{3\left(  1+\frac{1}{3}\left(  \frac{T}{T_{eff}%
}\right)  ^{4}\right)  }r_{1}}\sqrt[3]{1+\frac{9}{\,f\left(  T,d\right)
}\left(  1+\frac{1}{3}\left(  \frac{T}{T_{eff}}\right)  ^{4}\right)  }.
\end{equation}
Coming back to the redshift function, if we solve Eq.$\left(  \ref{PhiLTd}%
\right)  $ and we plug Eq.$\left(  \ref{oL}\right)  $ inside the solution, we
get%
\begin{gather}
\Phi\!\left(  r\right)  =-\frac{\sqrt{3}\,d^{4}}{2r_{0}\sqrt{r_{0}^{2}%
r_{1}^{2}\left(  T\right)  +4d^{4}}\,r_{1}\left(  T\right)  }\arctan\!\left(
\frac{\left(  2r+r_{0}\right)  r_{1}\left(  T\right)  }{\sqrt{3r_{0}^{2}%
r_{1}^{2}\left(  T\right)  +12d^{4}}}\right)  \nonumber\\
-\frac{3d^{4}\ln\!\left(  r_{1}^{2}\left(  T\right)  r^{2}+r_{0}r_{1}%
^{2}\left(  T\right)  r+r_{0}^{2}r_{1}^{2}\left(  T\right)  +3d^{4}\right)
}{4r_{0}^{2}r_{1}^{2}\left(  T\right)  }-\frac{\ln\!\left(  r\right)  }%
{2}+C.\label{Phi(r)LTd}%
\end{gather}
Note that $\Phi\!\left(  r\right)  \rightarrow-\infty$ when $r\rightarrow
+\infty$. However, we have to remember that there exists $\bar{r}$ such that
$b\left(  \bar{r}\right)  =0$. This means that we can choose $C$ in such a way
one can impose $\Phi\!\left(  \bar{r}\right)  =0$. To complete the evaluation
of this configuration, we need to compute $p_{t}\left(  d,T\right)  $.
Differently from the $T=0$ case, the form of the SET is not known. Nothing
forbids to assume that even at $T\neq0$, the SET can be traceless. Thus, one
gets%
\begin{equation}
\rho_{L,1}\left(  d,T\right)  -\omega_{L}\left(  d,T\right)  \rho_{L,1}\left(
d,T\right)  +2p_{t}\left(  d,T\right)  =0,
\end{equation}
which implies%
\begin{equation}  
p_{t}\left(  d,T\right)  =\rho_{L,1}\left(  d,T\right)  \frac{\left(
\omega_{L}\left(  d,T\right)  -1\right)  }{2}.
\end{equation}
$p_{t}\left(  d,T\right)  $ is in agreement with the $T=0$ case. On the other
hand, from the third EFE one finds that
\begin{equation}
p_{t}\left(  d,T\right)  =-\omega_{t}\left(  r,d,T\right)  \frac{r_{1}%
^{2}\left(  T\right)  \,}{\kappa d^{4}}=\omega_{t}\left(  r,d,T\right)
\frac{\rho_{L,1}\left(  d,T\right)  \,}{\kappa},
\end{equation}
where%
\begin{equation}
\omega_{t}\left(  r,d,T\right)  =-\frac{\left(  3r_{1}^{2}\left(  T\right)
\omega_{L}^{2}r^{3}+3r_{1}^{2}\left(  T\right)  \omega_{L}r_{0}^{3}%
-12d^{4}\omega_{L}r+9d^{4}\omega_{L}r_{0}+r_{1}^{2}\left(  T\right)
r^{3}-r_{1}^{2}\left(  T\right)  r_{0}^{3}-3r_{0}d^{4}\right)  }{4\left(
r_{1}^{2}\left(  T\right)  r^{3}-r_{1}^{2}\left(  T\right)  r_{0}%
^{3}+3r\,d^{4}-3r_{0}d^{4}\right)  }.
\end{equation}
It is clear that we have the same problem appeared in previous work on zero temperature Casimir wormholes \cite{CW},  . Therefore
the final SET in presence of temperature will be written in the following way%
\begin{gather}
T_{\mu\nu}=\frac{\rho_{L,1}\left(  d,T\right)  \,}{\kappa}\left[  diag\left(
-1,-\omega_{L}\left(  d,T\right)  ,\frac{\left(  \omega_{L}\left(  d,T\right)
-1\right)  }{2},\frac{\left(  \omega_{L}\left(  d,T\right)  -1\right)  }%
{2}\right)  \right.  \nonumber\\
+\left.  \left(  \omega_{t}\left(  r,d,T\right)  -\frac{\left(  \omega
_{L}\left(  d,T\right)  -1\right)  }{2}\right)  diag\left(  0,0,1,1\right)
\right]  ,\label{TmnL}%
\end{gather}
where the first term of $T_{\mu\nu}$ is traceless.

\subsection{Variable Plates distance separation at Low Temperature}
\label{p2B}
In this subesction we consider the energy density $\left(  \ref{rhoL2}\right)
$. The first modification we have to consider is%
\begin{equation}
T_{eff}=\frac{\hbar c}{2dk_{B}}\longrightarrow\frac{\hbar c}{2rk_{B}}.
\end{equation}
However, in order to keep the ratio $T/T_{eff}$, it is convenient to adopt
this change of scale%
\begin{equation}
\frac{\hbar c}{2dk_{B}}=\frac{\hbar c}{2dk_{B}}\frac{d}{r}=T_{eff}\frac{d}{r}.
\end{equation}
Then $f\left(  T,d\right)  $ becomes%
\begin{equation}
f\left(  T,r\right)  =1+\frac{45\zeta\left(  3\right)  }{\pi^{3}}\left(
\frac{T}{T_{eff}}\frac{r}{d}\right)  ^{3}-\left(  \frac{T}{T_{eff}}\frac{r}%
{d}\right)  ^{4}.
\end{equation}
It is immediate to see that we cannot include $f\left(  T,r\right)  $ as it
is, but we have to consider term by term. This means that the energy density
$\left(  \ref{rhoL2}\right)  $ can be written as%
\begin{equation}
\rho_{L,2}\left(  r,T\right)  =\hbar c\left[  -\frac{\pi^{2}}{720r^{4}}%
-\frac{\zeta\left(  3\right)  }{16d^{3}\pi r}\left(  \frac{T}{T_{eff}}\right)
^{3}+\frac{\pi^{2}}{720d^{4}}\left(  \frac{T}{T_{eff}}\right)  ^{4}\right]  .
\end{equation}
Despite of the complicated expression, the first EFE can be solved
\begin{align}
b\left(  r\right)   &  =r_{0}-\frac{8\pi G}{c^{4}}\int_{r_{0}}^{r}\rho
_{L,2}\left(  r^{\prime},T\right)  r^{\prime2}dr^{\prime}\nonumber\\
&  =r_{0}-\frac{\pi^{3}}{90}\ell_{P}^{2}\left[  \left(  \frac{1}{r}-\frac
{1}{r_{0}}\right)  +\ln\left(  \frac{r}{r_{0}}\right)  \frac{45\zeta\left(
3\right)  }{\pi^{3}d^{3}}\left(  \frac{T}{T_{eff}}\right)  ^{3}-\frac{1}%
{d^{4}}\left(  r^{3}-r_{0}^{3}\right)  \left(  \frac{T}{T_{eff}}\right)
^{4}\right]  .
\end{align}
As we can see, the only convergent term corresponds to the $T=0$ Casimir
energy density which has been found in previous work on zero temperature Casimir wormholes \cite{CW}. The last term is such
that the following property of a Traversable wormhole  $1-b(r)/r>0$ is violated. Therefore this
case will be discarded.

\section{High Temperature corrections to the Casimir Wormhole}

\label{p3}

 For the high temperature case, namely $ T\gg T_{eff} $, the thermal Casimir effect predicts the following corrections%
\begin{equation}
\qquad E\left(  d,T\right)  =-\frac{k_{B}TS}{8\pi d^{2}}\left[  \zeta\left(
3\right)  +\left(  \frac{4\pi T}{T_{eff}}+2\right)  \exp\left(  -\frac{2\pi
T}{T_{eff}}\right)  +O\left(  \exp\left(  -\frac{4\pi T}{T_{eff}}\right)
\right)  \right]  , \label{EHigh}%
\end{equation}
where $T_{eff}$ has been defined in Eq. $\left(  \ref{Teff}\right)  $. Its related pressure is defined by%
\begin{equation}
P\left(  d,T\right)  =-\frac{k_{B}T}{4\pi d^{3}}\zeta\left(  3\right)  . \label{PH}%
\end{equation}
$\zeta\left(  x\right)  $ is the Riemann zeta function. Following Ref.\cite{CW} we can consider the plates positioned at a
distance either parametrically fixed or radially varying. 
For high temperature corrections, it is reasonable to take into
account only the leading order since the other terms are exponentially
suppressed. Therefore, from Eq.$\left(  \ref{EHigh}\right)  $ we take%
\begin{equation}
\qquad E\left(  d,T\right)  =-\frac{k_{B}TS}{8\pi d^{2}}\zeta\left(  3\right)
.
\end{equation}
Like in the low temperature approximation, we can consider two possible configurations:

\begin{enumerate}
\item We divide $E\left(  d,T\right)  $ with a volume term of the form $V=Sd$,
leading to the following form of the energy density%
\begin{equation}
\rho_{H,1}\left(  d,T\right)  =-\frac{k_{B}T}{8\pi d^{3}}\zeta\left(
3\right)  .\label{rho1}%
\end{equation}
The related pressure is represented by Eq.$\left(  \ref{PH}\right)  $ leading
to the following EoS%
\begin{equation}
\frac{P\left(  d,T\right)  }{\rho_{H,1}\left(  d,T\right)  }=\omega
_{H}=2.\label{omegaH}%
\end{equation}
Note the difference with respect to the $T=0$ case where $\omega=3$.

\item In the same spirit of previous work on zero temperature Casimir wormholes \cite{CW}, we promote the plates separation
distance $d$ to a radial variable $r$ and we divide $E\left(  r,T\right)  $
with a volume term of the form $V=Sr$, leading to the following form of the
energy density%
\begin{equation}
\rho_{H,2}\left(  r,T\right)  =-\frac{k_{B}T}{8\pi r^{3}}\zeta\left(
3\right)  .\label{rho2}%
\end{equation}
Even in this case, we have that the EoS leads to the same value of Eq.$\left(
\ref{omegaH}\right)  $, namely $\omega_{H}=2$.
\end{enumerate}

We begin to consider the case 1)

\subsection{Constant Plates separation}

In this section we consider the profile $\left(  \ref{rho1}\right)  $ which
produces a shape function of the form%
\begin{equation}
b\left(  r\right)  =r_{0}-\frac{8\pi G}{c^{4}}\left(  \frac{k_{B}T}{8\pi
d^{3}3}\right)  \zeta\left(  3\right)  \left(  r^{3}-r_{0}^{3}\right)
=r_{0}-\frac{l_{P}^{2}\zeta\left(  3\right)  }{6d^{4}}\left(  \frac{T}%
{T_{eff}}\right)  \left(  r^{3}-r_{0}^{3}\right)  .\label{b(r)C}%
\end{equation}
Except for the coefficient in front of the cubic term which is different, the
shape function is formally the same of Eq.$\left(  \ref{b(r)LTd}\right)  $.
This means that, even in this case, the flare-out condition%
\begin{equation}
b^{\prime}\left(  r_{0}\right)  =-\frac{l_{P}^{2}\zeta\left(  3\right)
}{2d^{4}}\left(  \frac{T}{T_{eff}}\right)  r_{0}^{2}<1
\end{equation}
is always satisfied.  Even in this case by looking at its analytic form,
we can claim that there exists an $\bar{r}$ such that $b\left(  \bar
{r}\right)  =0$. Indeed, one finds%
\begin{equation}
\bar{r}=r_{0}\sqrt[3]{1+\frac{l_{1}^{2}\left(  d,T\right)  }{r_{0}^{2}}%
}\label{rbar}%
\end{equation}
where we have defined%
\begin{equation}
l_{1}\left(  d,T\right)  =\frac{d^{2}}{l_{P}}\sqrt{\frac{6}{\zeta\left(
3\right)  }\left(  \frac{T_{eff}}{T}\right)  }\simeq\frac{5\times10^{24}%
}{\sqrt{T}}m\label{l(a,T)}%
\end{equation}
and where we have considered a plates separation of the order of
$d\simeq10^{-6}m$ and $T_{eff}$ $\simeq5\times10^{2}\ ^{\circ}K$. Note that
for very high $T$, $l\left(  d,T\right)  \rightarrow0$. Plugging Eq.$\left(
\ref{b(r)C}\right)  $ into Eq.$\left(  \ref{pr}\right)  $, one finds%
\begin{equation}
\Phi'\!\left(  r\right)  =\frac{r_{0}^{3}+r_{0}l_{1}^{2}\left(
d,T\right)  -\left(  3\omega
_{H}+1\right)  r^{3}}{2r\left(  r\left(  r^{2}%
+l_{1}^{2}\left(  d,T\right)  \right)  -r_{0}\left(  l_{1}^{2}\left(
d,T\right)  +r_{0}^{2}\right)  \right)  }.\label{Phi'(r)H}%
\end{equation}
Close to the throat, we obtain%
\begin{equation}
\Phi'\!\left(  r\right)  \simeq\frac{l_{1}^{2}\left(  d,T\right)
-3\omega r_{0}^{2}}{2\left(  l_{1}^{2}\left(  d,T\right)  +r_{0}^{2}\right)
\left(  r-r_{0}\right)  }.
\end{equation}
Therefore, if we choose%
\begin{equation}
\omega
_{H}=\frac{l_{1}^{2}\left(  d,T\right)  }{3r_{0}^{2}},\label{oHH}%
\end{equation}
the horizon is not formed. Moreover, from Eq.$\left(  \ref{omegaH}\right)  $,
we can determine the size of the throat. Indeed, we find%
\begin{equation}
r_{0}=\frac{\sqrt{6}l_{1}\left(  d,T\right)  }{6}\simeq\frac{5.6\times10^{19}%
}{\sqrt{T}}m.
\end{equation}
Thus, the effect of high temperature corrections on a Casimir energy is a
reduction of the throat size. Nevertheless, we have to observe that in a
laboratory $T\simeq10^{8}\
{{}^\circ}%
K$ as an ideal result and this means that $r_{0}\simeq10^{11}m$ which is an
order of magnitude bigger than the solar system size. Anyway to understand if a TW can be supported by a hot Casimir source, we have to solve Eq.$\left(
\ref{Phi'(r)H}\right)  $ with the condition $\left(  \ref{oHH}\right)  $. One
finds%
\begin{equation}
\Phi\!\left(  r\right)  =-\frac{\ln\!\left(  r\right)  }{2}-\frac{l_{1}%
^{2}\left(  d,T\right)  }{4r_{0}^{2}}\ln\!\left(  l_{1}^{2}\left(  d,T\right)
+r^{2}+rr_{0}+r_{0}^{2}\right)  -\frac{l_{1}^{2}\left(  d,T\right)  }%
{2r_{0}\sqrt{4l_{1}^{2}\left(  d,T\right)  +3r_{0}^{2}}}\,\arctan\!\left(
\frac{2r+r_{0}}{\sqrt{4l_{1}^{2}\left(  d,T\right)  +3r_{0}^{2}}}\right)  +C.
\end{equation}
The redshift function is divergent for $r\rightarrow\infty$. However, with the help of Eq.$\left(  \ref{rbar}\right)  $, nothing
forbids to cure such a divergence by imposing that $\Phi\!\left(  \bar
{r}\right)  =0$. Then,
one gets%
\begin{gather}
\Phi\!\left(  r\right)  =-\frac{1}{2}\ln\!\left(  \frac{r}{\bar{r}}\right)
-\frac{l_{1}^{2}\left(  d,T\right)  }{4r_{0}^{2}}\ln\!\left(  \frac{l_{1}%
^{2}\left(  d,T\right)  +r^{2}+rr_{0}+r_{0}^{2}}{l_{1}^{2}\left(  d,T\right)
+\bar{r}^{2}+\bar{r}r_{0}+r_{0}^{2}}\right)  \nonumber\\
-\frac{l_{1}^{2}\left(  d,T\right)  }{2r_{0}\sqrt{4l_{1}^{2}\left(
d,T\right)  +3r_{0}^{2}}}\left(  \,\arctan\!\left(  \frac{2r+r_{0}}%
{\sqrt{4l_{1}^{2}\left(  d,T\right)  +3r_{0}^{2}}}\right)  -\arctan\!\left(
\frac{2\bar{r}+r_{0}}{\sqrt{4l_{1}^{2}\left(  d,T\right)  +3r_{0}^{2}}%
}\right)  \right)  .
\end{gather}
It is interesting to observe that for $T\ggg T_{eff}$, the redshift function
reduces to%
\begin{equation}
\Phi\!\left(  r\right)  \simeq-\frac{1}{2}\ln\!\left(  \frac{r}{\bar{r}%
}\right)
\end{equation}
with $\bar{r}\rightarrow r_{0}$. This means that the set of definition of the
variable $r$ shrinks for both $\Phi\!\left(  r\right)  $ and $b\left(
r\right)  $. To complete the calculation we need to compute $p_{t}(r)$.
Nevertheless, before going on we can gain information on $p_{t}(r)$ by
assuming that even for high temperature the Casimir SET is traceless. This
means that%
\begin{equation}
p_{t}(d,T)=-\frac{1}{2}\rho_{H,1}\left(  d,T\right)
\end{equation}
and the corresponding SET can be written as%
\begin{equation}
T_{\mu\nu}=\rho_{H,1}\left(  d,T\right)  \left[  diag\left(
1,2,-1/2,-1/2\right)  \right]  .
\end{equation}
However, from the third EFE, one finds%
\begin{equation}
p_{t}(d,T)=-\omega_{t}^{H}\left(  r,d,T\right)  \frac{3\,}{\kappa l_{1}%
^{2}\left(  d,T\right)  }=\omega_{t}^{H}\left(  r,d,T\right)  \frac{\rho
_{H,1}\left(  d,T\right)  \,}{\kappa},
\end{equation}
where%
\begin{equation}
\omega_{t}^{H}\left(  r,d,T\right)  =-\frac{l_{1}^{4}\left(  d,T\right)
\left(  r^{2}+rr_{0}-3r_{0}^{2}\right)  +3r_{0}^{4}\left(  r^{2}+rr_{0}%
+r_{0}^{2}\right)  }{4r_{0}^{4}\left(  l_{1}^{2}\left(  d,T\right)
+r^{2}+rr_{0}+r_{0}^{2}\right)  }.
\end{equation}
Like for Eq.$\left(  \ref{TmnL}\right)  $, we can separate the SET into a
traceless term and a trace part. One finds%
\begin{gather}
T_{\mu\nu}=\frac{\rho_{H,1}\left(  d,T\right)  \,}{\kappa}\left[  diag\left(
1,2,-\frac{1}{2},-\frac{1}{2}\right)  \right.  \nonumber\\
+\left.  \left(  \omega_{t}^{H}\left(  r,d,T\right)  +\frac{1}{2}\right)
diag\left(  0,0,1,1\right)  \right]  ,
\end{gather}
where the first term of $T_{\mu\nu}$ is traceless.

\subsection{Variable Plates separation}

In this subsection we consider the energy density $\left(  \ref{rho2}\right)
,$ representing the Casimir source with high temperature corrections. The
first EFE leads to%
\begin{equation}
b\left(  r\right)  =r_{0}-\frac{8\pi G}{c^{4}}\left(  \frac{k_{B}T}{8\pi
}\right)  \zeta\left(  3\right)  \int_{r_{0}}^{r}\frac{dr^{\prime}}{r^{\prime
}}=r_{0}-l\left(  T\right)  \ln\left(  \frac{r}{r_{0}}\right)  ,\label{b(r)H}%
\end{equation}
where we have defined%
\begin{equation}
l\left(  T\right)  =\zeta\left(  3\right)  l_{P}^{2}\frac{k_{B}T}{\hbar c}.
\end{equation}
From the shape function $\left(  \ref{b(r)H}\right)  $, one finds that the
flare out condition is always satisfied since%
\begin{equation}
b^{\prime}\left(  r_{0}\right)  =-\frac{l\left(  T\right)  }{r_{0}}<1.
\end{equation}
Moreover from the condition $1-b\left(  r\right)  /r>0$, it is possible to
check that there exists $r=\bar{r}$ such that $b\left(  \bar{r}\right)  =0$,
where%
\begin{equation}
\bar{r}=r_{0}\exp\left(  \frac{r_{0}}{l\left(  T\right)  }\right)  .
\end{equation}
Finally, to obtain a TW we need to compute the redshift function. From
Eq.$\left(  \ref{pr}\right)  $, one finds%
\begin{equation}
\Phi^{\prime}\!\left(  r\right)  =-\frac{r_{0}-l\left(  T\right)  \left(
\omega+\ln\!\left(  r/r_{0}\right)  \right)  }{2r\left(  l\left(  T\right)
\left(  \ln\!\left(  r/r_{0}\right)  \right)  +\left(  r-r_{0}\right)
\right)  },\label{pr0a}%
\end{equation}
where we have imposed the EoS $p_{r}\left(  r\right)  =\omega\rho\left(
r\right)  $. Close to the throat, the r.h.s. of Eq.$\left(  \ref{pr0a}\right)
$ can be approximated by%
\begin{equation}
\Phi\!^{\prime}\left(  r\right)  \simeq\frac{r_{0}-\omega l\left(  T\right)
}{2r_{0}\left(  r-r_{0}\right)  }.
\end{equation}
Following Ref.\cite{CW} we can choose $\omega$ such that $\Phi^{\prime
}\!\left(  r\right)  =0$ avoiding the appearance of a horizon. This can be
done if%
\begin{equation}
\omega=\frac{r_{0}}{l\left(  T\right)  }.\label{oH}%
\end{equation}
Note that from Eq.$\left(  \ref{PH}\right)  $, it is possible to fix the value
of $\omega$. Indeed%
\begin{equation}
\frac{P\left(  r,T\right)  }{\rho_{H,2}\left(  r,T\right)  }=\omega
=2,\label{omH}%
\end{equation}
which is different form the zero temperature value $\omega=3$. From
Eq.$\left(  \ref{omH}\right)  $ and Eq.$\left(  \ref{oH}\right)  $, one finds%
\begin{equation}
r_{0}=2l\left(  T\right)  \simeq\frac{10^{-67}m}{K}T.
\end{equation}
It is immediate to see that not even with a temperature much larger than the
Planck Temperature%
\begin{equation}
T_{P}=\sqrt{\frac{\hbar c^{5}}{Gk_{B}^{2}}}\simeq1.416784\times10^{32}K,
\end{equation}
the solution has a physical meaning. Therefore this solution will not be
considered. However, another possibility comes from the following observation.
The energy density $\left(  \ref{rho2}\right)  $ can be rewritten in the
following way%
\begin{equation}
\rho_{H,2}\left(  r,T\right)  =-\frac{\hbar c}{16\pi r^{3}}\frac{\zeta\left(
3\right)  }{\lambda_{C}\left(  T\right)  },
\end{equation}
where we have introduced the Casimir thermal wave length%
\begin{equation}
\lambda_{C}\left(  T\right)  =\frac{\hbar c}{2k_{B}T}.
\end{equation}
In the high temperature approximation, or long distance approximation, the
following inequality is satisfied%
\begin{equation}
\frac{2\pi}{\lambda_{C}\left(  T\right)  }d\gg1,
\end{equation}
where $d$ is the plates separation. By promoting the $d$ distance to a
variable distance $r$, we can write%
\begin{equation}
\frac{1}{\lambda_{C}\left(  T\right)  }=\frac{A}{2\pi r},\qquad A\gg1.
\end{equation}
Thus $\rho_{H,2}\left(  r,T\right)  $ can be cast into the form
\begin{equation}
\rho_{H,2}\left(  r,T\right)  =-A\frac{\hbar c}{32\pi^{2}r^{4}}\zeta\left(
3\right)  .
\end{equation}
Thus the first EFE becomes%
\begin{equation}
b\left(  r\right)  =r_{0}-\frac{\hbar G}{4\pi c^{3}}A\zeta\left(  3\right)
\int_{r_{0}}^{r}\frac{dr^{\prime}}{r^{\prime2}}=r_{0}+r_{T}^{2}\left(
\frac{1}{r}-\frac{1}{r_{0}}\right)  ,
\end{equation}
where we have defined%
\begin{equation}
r_{T}^{2}=\frac{A}{4\pi}\zeta\left(  3\right)  l_{P}^{2}.
\end{equation}
As we can see, we have the same formal expression found in previous work on
zero temperature Casimir wormholes \cite{CW}. Therefore, if we solve the
second EFE $\left(  \ref{pr}\right)  $, we get%
\begin{equation}
\Phi^{\prime}\!\left(  r\right)  =\frac{\left(  \left(  -\omega+1\right)
r_{0}-r\right)  r_{T}^{2}+rr_{0}^{2}}{2r\left(  r-r_{0}\right)  \left(
rr_{0}+r_{T}^{2}\right)  }%
\end{equation}
and close to the throat we find that a horizon can be avoided if%
\begin{equation}
\omega=\frac{r_{0}^{2}}{r_{T}^{2}}.
\end{equation}
However, this time the value of $\omega$ is given by%
\begin{equation}
\frac{P\left(  r,T\right)  }{\rho_{H,2}\left(  r,T\right)  }=\omega=2
\end{equation}
and not $\omega=3$, like in previous work on zero temperature Casimir
wormholes \cite{CW}. Except for this point, we can now estimate the value of
the throat which is%
\begin{equation}
r_{0}=\sqrt{\frac{A}{2\pi}}l_{P}.
\end{equation}
Compared with the zero temperature result \cite{CW}, we can see that the
effect of the temperature is of enlarging the size of the throat.

\section{Conclusions}
\label{p4}
   In this paper, we have investigated how finite temperature
corrections to the Casimir Wormhole source \cite{CW} can modify the potential
traversability property. To do calculations in practice, we have considered
the low and high temperature approximations in two particular configurations
of the plates. In one configuration, the plates have been taken parametrically
fixed and in the other one they have been taken radially varying. When the
plates are parametrically fixed, for both low and high temperature
approximations, one finds the same prediction obtained also at zero
temperature \cite{GABTW}, namely the size of the throat is really huge. In
particular, for the low temperature, we obtain approximately the same result
of the zero temperature, while for the high temperature we have the
possibility of reducing the throat size because of the presence of the
$\sqrt{T}$ in the denominator. However, to obtain a physically relevant
result, one should be able to produce a temperature so high which is not
realizable with the currently technology. Indeed, the maximum realizable
temperature is of the order of $T\simeq10^{8}\
{{}^\circ}%
K$ and this predicts a Traversable Wormhole with a throat size bigger than the
solar system. On the other hand, when the plates are considered radially
varying when the low temperature is examined, one finds no solutions
compatible with the traversability features and for the high temperature
approximation, we need to consider a twofold approach. The first one deals
with the energy density as it is predicting a solution that, unfortunately is
not traversable either in principle. The second one introduces the Casimir
thermal wave length and, for this reason, one is forced to introduce also an
extra radial component. The final result looks like the original Casimir
wormhole, but with an additional large constant appearing in front of the
energy density which allows the wormhole throat to be no more Planckian.
Therefore, from one side, we have Planckian Traversable Wormholes which can
have the size of the throat slightly enlarged and from the other side we have
giant Traversable Wormholes which have a throat that cannot be considerably
reduced. Nevertheless, to summarize, we can certainly claim that the thermal
effects do not destroy the traversablity of the  \textquotedblleft\textit{Hot
Casimir Wormhole}\textquotedblright. To this purpose, it could be interesting
to consider quantum gravitational corrections to the original classical
background following the formalism outlined in Ref.
\cite{Remo,Remo1,RGFLCQG,RGFLPLB,RGFLPRD,RGFLBook}, where the graviton one
loop contribution to a classical energy in a traversable wormhole background
was computed. It will be also interesting to construct such hot traversable
wormholes using the gravitational Casimir effect \cite{j}. This can be done
for both general relativity and teleparallel gravity. In fact, we expect that
the Casimir effect be different for general relativity and teleparallel
gravity. This is because even though both teleparallel gravity and general
relativity are identical in the bulk of any spacetime manifold, they differ
from each other via a total derivative term, which can be converted into a
boundary term. Now as the Casimir effect is a boundary effect, the
gravitational Casimir effect will be different for general relativity and
teleparallel gravity \cite{j1}. This in turn can be used to produce novel
solutions involving traversable wormholes. Thus, we expect that traversable
wormholes produced by general relativity and teleparallel gravity will be
different from each other. Apart from gravitational Casimir wormholes in
general relativity and teleparallel gravity, it will be interesting to
generalize these results to $f(R)$ and $f(T)$ gravity. These results can also
be generalized to other modified theories of gravity. As the gravitational
Casimir effect will directly be modified in any theory of modified gravity, it
is expected that non-trivial Casimir wormhole solutions can be constructed
using such modified theories of gravity.

It is possible to study the Casimir effect using holography. Thus, holographic
dual to the Casimir effect in the boundary conformal field theory has been
studied using a suitable dual geometry \cite{du}. The holography has been used
to find a general relation between Casimir effect and the Weyl anomaly. It was
also observed that the relation between them is universal. In fact, they are
fixed by the central charge of the boundary conformal field theory. It would
be interesting to investigate the thermal generalization of these results.
Thus, we can use a thermal conformal field theory, and find the suitable
geometry dual to it. Then we can use this duality to find the holographic dual
to a thermal Casimir effect. It would be interesting to use such results to
analyze the holographic dual to various different conformal field theories,
using suitable geometries dual such conformal field theories.

\end{document}